\documentclass[prl,aps,twocolumn,superscriptaddress,amsmath,amssymb]{revtex4}
\usepackage{amsmath,epsfig}
\usepackage{color}
\definecolor{Red}{rgb}{1,0,0}

\newcommand{\ket}[1]{|#1\rangle}

\begin{document}
%
%
%
%
\title{Charge frustration and quantum criticality for strongly
  correlated fermions}

\author{Liza Huijse}
\affiliation{Institute for Theoretical Physics, University of Amsterdam,
Valckenierstraat 65, 1018 XE Amsterdam, The Netherlands}

\author{James Halverson}
\affiliation{Department of Physics, University of Virginia,
Charlottesville, VA 22904-4714 USA}

\author{Paul Fendley}
\affiliation{Department of Physics, University of Virginia,
Charlottesville, VA 22904-4714 USA}

\author{Kareljan Schoutens}
\affiliation{Institute for Theoretical Physics, University of Amsterdam,
Valckenierstraat 65, 1018 XE Amsterdam, The Netherlands}

\date{April 1, 2008,  revised: October 08, 2008}

\begin{abstract}

We study a model of strongly correlated electrons on the square
lattice which exhibits charge frustration and quantum critical
behavior. The potential is tuned to make the interactions
supersymmetric. We establish a rigorous mathematical result which relates quantum ground states
to certain tiling configurations on the square lattice. For periodic boundary conditions 
this relation implies that the number of ground
states grows exponentially with the {\em linear} dimensions of the
system. We present substantial analytic and numerical evidence that for open boundary
conditions the system has gapless edge modes.

\end{abstract}

\pacs{PACS numbers: 71.27.+a, 05.30.-d, 11.30.Pb}

\maketitle

Quantum criticality is a key notion in the analysis of non-Fermi 
liquid behavior of strongly correlated electrons. 
In general, quantum criticality is expected when 
parameters are tuned such that competing orders are perfectly balanced
\cite{sach}. In two or three spatial dimensions, the identification of quantum critical
behavior is quite difficult, due to a lack of sufficiently
powerful analytic methods and to the limited reach of numerics.

In this Letter, we identify quantum critical behavior in a specific
model for spinless itinerant fermions on a two-dimensional square 
lattice. The Hamiltonian combines standard hopping terms with 
strong repulsive interactions. At filling $f=1/2$ (one fermion
per two sites) this model develops a Mott insulating phase with
a fermion on every other lattice site. In \cite{ZH}, it was shown that a 
related model exhibits a stripe phase upon modest doping. 
Here we study the model in the `strongly doped Mott insulator phase' at intermediate density.

In the model at hand, interactions have been tuned to obtain 
a property called supersymmetry, which we employed in our analysis.
In previous work (for a review see \cite{HS}), we demonstrated that this model 
develops a form of quantum charge
frustration that leads to an {\em extensive ground
state entropy} \cite{FS,vE}. Here we show 
that this same model on the square lattice with an edge possesses quantum critical edge excitations.
We find that this quantum criticality arises due to the competition 
of two forms of charge order at densities $f=1/5$ 
and $f=1/4$.

Our degrees of freedom are spinless fermions living on the square lattice.
A fermion at site $i$ is created
by the operator $c_i^\dagger$ with $\{c_i,c^\dagger_j\}=\delta_{ij}$.
The fermions have a {\it hard core}, meaning that they are not only forbidden to be
on the same site as required by Fermi statistics, but are also
forbidden to be on adjacent sites. Their creation operator is
$d_i^\dagger=c_i^\dagger {\cal P}_{<i>}$, where 
\begin{equation}
{\cal P}_{<i>}= \prod_{j\hbox{ next to } i} (1-c^\dagger_j c_j) \nonumber
\label{proj}
\end{equation}
is zero if any site next to $i$ is occupied.

Our model, first introduced in \cite{FSd}, is easiest to define in terms of the
``supersymmetry'' operator $Q=\sum_i d^\dagger_i$.  The Hamiltonian is
\begin{equation}
H=\{Q,Q^\dagger\}= \sum_{<ij>} d^\dagger_i d_j +  \sum_i {\cal P}_{<i>}.\nonumber
\label{ham}
\end{equation}
The model is supersymmetric because $Q^2=(Q^\dagger)^2=0$, which then implies
that $[Q,H]=[Q^\dagger,H]=0$.
The latter term in the Hamiltonian combines a chemical potential and a repulsive 
next-nearest-neighbor potential. On the square lattice it reads
\begin{equation}
\sum_i {\cal P}_{<i>}=N-4F+\sum_{i}{V}_{<i>}\nonumber
\label{potential}
\end{equation}
where ${V}_{<i>}+1$ is the number of particles adjacent to $i$, unless
there are none, in which case ${V}_{<i>}=0$. $N$ is the number of sites. The 
operator $F=\sum_i d^\dagger_i d_i$ counts the number of fermions and commutes with the Hamiltonian. The chemical
potential is fixed at $\mu=4$ and there is no {\em a priori} constraint on the
particle number.

The problem of counting the ground states in this supersymmetric
lattice model turns out to be related to some simple-to-describe
(but often difficult to solve) geometrical counting problems (see
\cite{FSd,FS}). A heuristic way of understanding this is from the ``3-rule'': to
minimize the energy, fermions prefer to be mostly 3 sites apart (with details 
depending on the lattice). Using techniques
from cohomology, it was proved for several
lattices that the configurations which satisfy the 3-rule are in one-to-one 
correspondence with ground states \cite{FS}.

The correspondence between these configurations and ground states holds
for a wide variety of boundary conditions. This indicates that the
ground state wavefunctions
exhibit a form of charge ordering. This ordering is akin to the N\'eel
ordering in an antiferromagnet: the ordered state is not an exact
eigenstate, but one can find a non-vanishing antiferromagnetic order
parameter for the ground state. In this sense, the ordered state
dominates the ground state. For our model, the structure is even richer, because there are typically an
exponentially growing number of ways of satisfying the
3-rule. Such behavior is what we mean by {\em charge frustration}. On generic
lattices, the number of ground states grows exponentially with the
full (two-dimensional) volume of the system \cite{FSd,vE}, see also \cite{eng}. 
However, for the square lattice described here,
we will show that for periodic boundary conditions the number of ground states grows exponentially with the {\em linear}
dimensions of the system.

An important object in supersymmetric theories
is the {\em Witten index} $W$, which is defined as $W=N_b-N_f$, where
$N_b$ is the number of bosonic ground states (those with an even
number of fermions), and $N_f$ is the number of fermionic ground
states (those with an odd number of fermions) \cite{Witten}. By
definition, $W$ is a lower bound on the number of ground states. 

\begin{figure}[h!]
\begin{center}
\includegraphics[width= .45\textwidth]{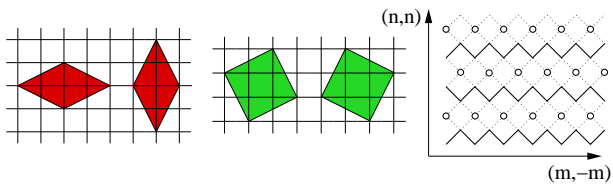}
\caption{The diamonds on the left and squares in the middle. On the right we show the square lattice rotated
by $45^{\circ}$. Sublattice $S_1$ is indicated by the circles and
sublattice $S_2$ is indicated by the drawn lines.}
\label{fig:rhombuses}
\end{center}
\end{figure}

We show in this paper that for the square lattice, the ground states are in one-to-one correspondence with 
configurations that come from tiling the plane using the
four rhombuses illustrated in fig. \ref{fig:rhombuses}, which
for obvious reasons we call diamonds and squares. Here the particles sit at the corners of the rhombuses,
along the edges they are 3 sites apart. A key result was proved rigorously by Jonsson 
\cite{Jonsson}. He showed that for the square lattice with periodic boundary conditions
in both directions, the Witten index is simply related to tilings with these rhombuses. Precisely, let $t_b$ ($t_f$) be 
the number of ways of tiling the torus with these four rhombuses, so that
there are an even (odd) number of fermions [or equivalently, an even (odd) number
of rhombuses]. The most general version of this theorem
allows for any type of torus on the square lattice with periodicities $\vec{u}=(u_1,u_2)$
and $\vec{v}=(v_1,v_2)$. The expression for the Witten index then reads \cite{Jonsson}
\begin{equation}
W_{u,v} = N_b - N_f = t_b - t_f - (-1)^{d_-} \theta_{d_-} \theta_{d_+},\nonumber
\end{equation}
where $d_{\pm} \equiv \hbox{gcd}(u_1 \pm u_2,v_1 \pm v_2)$ and 
\begin{equation}
\theta_d \equiv \left\{ \begin{array}{ll}
2 & \textrm{if $d=3k$, with $k$ integer}\\
-1 & \textrm{otherwise.}
\end{array} \right. \nonumber
\end{equation}

A natural extension of Jonsson's theorem is to relate the {\em total} number of ground
states to rhombus tilings:
\begin{equation}
N_b + N_f = t_b + t_f+ \Delta,
\label{PF}
\end{equation}
where $|\Delta|=|\theta_{d_-} \theta_{d_+}|$. When $\vec{u}=(m,-m)$ and $v_1+v_2=3p$ we
prove this relation explicitly with $\Delta=-(-1)^{(\theta_m+1)p} \theta_{d_-} \theta_{d_+}$ (see below).

In our supersymmetric models, the problem of computing the total number of ground states
reduces to finding the cohomology of the supercharge $Q$ \cite{FSd}. More precisely, the dimension of the non-trivial
cohomology of $Q$ corresponds to the total number of linearly independent ground states. The cohomology is the vector
space of states which are annihilated by $Q$ but which are not $Q$ of something else.

To compute the cohomology is in general very difficult. A useful theorem is the
`tic-tac-toe' lemma of \cite{botttu}. This says that under certain conditions,
the cohomology $H_Q$ for $Q = Q_1 + Q_2$ is the same as the cohomology
of $Q_1$ acting on the cohomology of $Q_2$. In an equation,
$H_Q = H_{Q_1}(H_{Q_2} ) \equiv H_{12}$, where
$Q_1$ and $Q_2$ act on different sublattices $S_1$ and
$S_2$.

The crucial step is to choose the right sublattices: for the square lattice we take a set of
disconnected sites for $S_1$ and a set of (periodic) chains for $S_2$ (see fig.
\ref{fig:rhombuses}). For free boundary conditions in either one or both of the diagonal directions 
[$(m,-m)$ and $(n,n)$] the full cohomology problem has been solved using
Morse theory \cite{BM}. These cases can also be solved using `tic-tac-toe' with $S_1$ and
$S_2$ as in fig. \ref{fig:rhombuses}. To solve $H_{Q_2}$, we start from the bottom chain. If a site on $S_1$ 
directly above this chain is occupied,
there is an isolated site on the bottom chain, leading to a vanishing $H_{Q_2}$. It follows that all sites 
directly above the bottom chain must be empty. Continuing this argument
for subsequent chains, one finds that all sites on $S_1$ must be empty. Computing
$H_{12}$ is now a trivial step. The dimension of $H_Q$ is related to the number of ground states on
the chains that constitute $S_2$. This depends on the exact boundary conditions. One finds that the
dimension of $H_Q$ is either one or zero, except for the
cylindrical case periodic in the $(m,-m)$-direction with $m=3p$ and $n=3q$ or $n=3q+1$. In that case the 
dimension of $H_Q$ is $2^K$, with $K$ the nearest integer to $2n/3$.

For periodic boundary conditions in both diagonal directions, solving $H_Q$ becomes much more involved. This
is due to the fact that in this case solving $H_{Q_2}$ does not
imply that all sites on $S_1$ are empty. Instead, there are many allowed configurations on $S_1$, and solving
$H_{12}$ becomes highly non-trivial. We tackled this problem in two steps. First, we
solved $H_{12}$ for the case where $S_2$ consists of 1 chain of arbitrary length, that is
$\vec{u}=(m,-m)$ and $\vec{v}=(1,2)$. Then, we extended this to the case where $S_2$ consists of an
arbitrary number of chains, that is $v_1+v_2=3p$. This establishes (\ref{PF}) with $\Delta$ as given above. 
Our proof, which will be presented elsewhere, resolves the ground state counting
problem of this highly frustrated system on the 2D square lattice. Counting rhombus
tilings, which is relatively easy \cite{Jonsson}, shows that indeed the number of ground states
grows exponentially with the linear dimensions of the system.

For $\vec{v}=(1,2)$, the tilings reduce to a linear sequence of tiles in arbitrary order. There are 
zero-energy ground states 
at all rational filling fractions in the range between 1/5 and 1/4. For
example, for a system with 30 sites there are 5 linearly independent ground
states with 6 particles and 92 with 7 particles. The tilings not only count, but indeed
seem to dominate the actual ground states. For specific lengths of the $S_2$ chain, we have seen numerically that the charge distribution of a ground state 
largely overlaps with that of the corresponding tiling. We can exploit this to 
gain physical insight. There are two uniform phases: all squares at 1/5 filling and all diamonds at 1/4 filling.
One diamond in a phase with all squares is then a zero-energy defect with fractional
charge 1/5. From counting the number of tilings with one such defect
it follows that a defect can have any momentum. This is reminiscent of a flat band. Flat bands usually arise
either from tuning the hopping terms of non-interacting particles on an exotic lattice,
or from tuning potential terms for strongly
interacting particles with negligible kinetic energy \cite{flat}. Here, however, the flat band arises from tuning the
potentials for particles with kinetic and potential energies of comparable size. Note
that filling this flat band with defects is slightly subtle. Since defects cannot sit on top of each other
and span over 4 sites, they obey a certain exclusion statistics.

\begin{figure}[h!]
\begin{center}
\includegraphics[width= .33\textwidth]{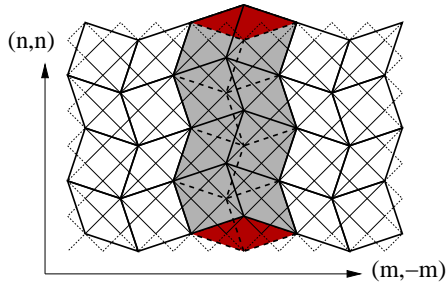}
\caption{Edge modes.}
\label{fig:egdemodes}
\end{center}
\end{figure}

We can exploit the effective geometric description of the space of ground states even more by comparing
periodic and free boundary conditions. For the square lattice wrapped around the torus with $\vec{u}=(m,-m)$ and $\vec{v}=(n,n)$
the number of ground states grows as $2^{2(n+m)/3}$
\cite{Jonsson}. On the cylinder, however, if one cuts the torus open  along the $\vec{u}$-direction
only $2^{2n/3}$ ground states remain. Finally, if one also cuts the cylinder open along the $\vec{v}$-direction
one is left with a unique ground state on the plane. What happens to this vast
number of ground states that disappear upon changing the boundary conditions? Consider the picture in fig.
\ref{fig:egdemodes}. If one identifies the dotted, zigzagged boundaries both in the $\vec{u}$ as well as in
the $\vec{v}$-direction, one finds that both the tiling with the drawn lines as well as the one with
the dashed lines represent ground states. However, if one only identifies the
boundaries in the $\vec{u}$-direction, then the tiling with the drawn lines no longer represents a ground
state. Instead, it has
two defects at the edges, which can propagate along the edge. The only
available scale for the energy of the edge mode is one over
the length of the edge, which suggests that the edge modes are gapless.

To further investigate criticality in this system, we have studied the model numerically 
with periodic boundary conditions
such that $\vec{u}=(L,0)$ and $\vec{v}$ equals $(0,2)$, $(1,2)$ or
$(3,3)$. In all three cases, we have found compelling evidence for
critical modes. We investigated how the spectrum changes upon
twisting the boundary condition along the $(L,0)$-direction from
periodic to anti-periodic. This is a powerful way of
distinguishing between critical and gapped states: for a gapped
state, the correlation length is finite and a change in the
boundary conditions will have an exponentially small effect on the
energy. In contrast, for a critical state the change in the energy
will be substantial since the correlation length goes to infinity.
More specifically, we will see that the energy has a parabolic
dependence on the boundary twist.

In addition to this, one can extract more quantitative properties from twisting the boundary condition.
For a critical supersymmetric system in 1D with a Fermi surface, we expect that its continuum limit is
described by an ${\cal N}$=2 superconformal field theory (SCFT). In such a theory, twisting the boundary
condition corresponds to going from the Ramond to the Neveu-Schwarz sector.
The twist can be carried out continuously and leads to a spectral flow
\cite{SS}. If we define the twist parameter $\alpha$ to be integer in the Ramond sector and
half-integer in the NS sector, the energy is a parabolic function of $\alpha$,
\begin{eqnarray}\label{spflow}
E_{\alpha} &=& E_{\alpha=0} - \alpha Q_{\alpha=0} +\alpha^2 c/3,
\end{eqnarray}
where $c$ is the central charge. $Q_{\alpha}$ depends linearly on $\alpha$ and is the sum of the left- and 
right-moving U(1) charges \cite{SS,BFK}. Their difference is conserved under 
the twist and is related to the fermion number.
In the lattice model, we can go from periodic to anti-periodic boundary conditions continuously by replacing the term that
hops a particle over the boundary $c^{\dag}_N c_1+$ h.c. by
$e^{2 \pi \imath \alpha}c^{\dag}_N c_1+$ h.c. The eigenvalues of the translation operator $p_{\alpha}$ will now depend
linearly on the twist parameter:
 \begin{eqnarray}
T_{\alpha}^L \ket{\psi} &=& e^{2 \pi \imath p_0 L} e^{2 \pi \imath \alpha F} \ket{\psi} \equiv e^{2 \pi
\imath p_{\alpha} L}\ket{\psi},
\end{eqnarray}
so $p_{\alpha}= p_0 + \alpha F/L$ where $L$ is the length of the
system and $F$ is the total number of particles in the state $\ket{\psi}$.

\begin{figure}[h!]
\begin{center}
\includegraphics[width= .40\textwidth]{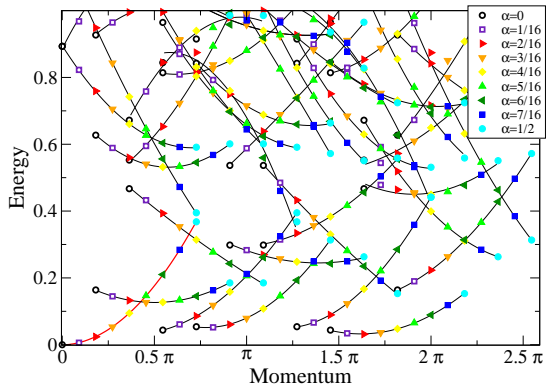}
\caption{In the plot we show nine energy spectra as a function of momentum for increasing values of the
twist parameter ($0\leq \alpha \leq 1/2$, with steps of $1/16$) for a system with 33 sites
($\vec{u}=(11,0)$ and $\vec{v}=(3,3)$) and 8 fermions. The lines are parabolic fits to the
numerical data. The results from the spectral flow analysis for the red line through the origin appear in red in table
\ref{tab:eoverc}.}
\label{fig:spflowplot}
\end{center}
\end{figure}

We computed the spectrum for various values of the twist parameter via exact diagonalization. We find that
the majority of states have a parabolic dependence on the twist parameter by fitting a parabola
to the energy levels as a function of the twist parameter, or equivalently, as a function of $p_{\alpha}$.
This clearly indicates that the system
is critical. An example is shown in fig. \ref{fig:spflowplot}.

For a critical system, the energy of the SCFT in (\ref{spflow}) is related to the numerically obtained value
of the energy via $E_{\textrm{num}}=2 \pi E_{\textrm{CFT}}v_F/L$, where $v_F$ is the Fermi velocity and $L$ the
system size. So by comparing the parabolic fit to the numerics with equation (\ref{spflow}), we
can obtain the ratios $E_{\alpha}/c$ and $Q_{\alpha}/c$.

\begin{table}[h!]
\caption{Results from spectral flow analysis for three types of tori $(L,0)\times\vec{v}$.
Here, $N$ denotes the number of
sites and $F$ the number of fermions. We show the results for the lowest energy level for each system.
The values for $E$ and $Q$ are given in the NS sector ($\alpha=1/2$) and $c$ is the central
charge. For the level in red boldface the
results are extracted from the red line in fig. \ref{fig:spflowplot}.} 
\begin{tabular}{r @{\hspace{1.1cm}} c @{\hspace{1.1cm}} r @{\hspace{1.1cm}} r @{\hspace{1.1cm}} r}
\hline
\hline
$N$ & $\vec{v}$ & $F$ & $E/c$ & $Q/c$ \\
\hline
18& $(3,3)$ &4&-0.0851&0.004 \\
36&$(3,3)$&8&-0.0841&-0.002 \\
15&$(3,3)$&4&0.0898&0.349 \\
21&$(3,3)$&4&0.0850&0.337 \\
24&$(3,3)$&5&0.0850&0.337 \\
30&$(3,3)$&7&0.0853&0.338 \\
\textcolor{Red}{\textbf{33}}&\textcolor{Red}{\textbf{(3,3)}}&\textcolor{Red}{\textbf{8}}&\textcolor{Red}{\textbf{0.0855}}&\textcolor{Red}{\textbf{0.338}}\\
9& $(1,2)$ &2&-0.0858&-0.005\\
18&$(1,2)$&4&-0.0842&-0.002\\
27&$(1,2)$&6&-0.0839&-0.001\\
17&$(1,2)$&4&0.0844&0.336\\
26&$(1,2)$&6&0.0840&0.335\\
35&$(1,2)$&8&0.0839&0.335\\
14&$(1,2)$&3&0.2666&0.701\\
23&$(1,2)$&5&0.2458&0.657\\
32&$(1,2)$&7&0.2432&0.652\\
16& $(0,2)$ &4&-0.0897&-0.014\\
24&$(0,2)$&6&-0.0889&-0.012\\
32&$(0,2)$&8&-0.0885&-0.011\\
12&$(0,2)$&3&0.0911&0.350\\
20&$(0,2)$&5&0.0900&0.348\\
28&$(0,2)$&7&0.0894&0.347\\
14&$(0,2)$&4&0.0855&0.338\\
22&$(0,2)$&6&0.0849&0.337\\
30&$(0,2)$&8&0.0847&0.336\\
\hline
\hline
\end{tabular}
\label{tab:eoverc}
\end{table}
We extracted values for $E_{\alpha}/c$ and $Q_{\alpha}/c$ via the above described method for three models
with up to 36 sites (see table \ref{tab:eoverc}). For the lowest energy levels, we typically find that 
$(E/c,Q/c)$ in the NS sector is either $(-1/12,0)$, $(1/12,1/3)$ or $(1/4,2/3)$, all with an accuracy of 
within $10\%$. These values occur in the Kac table for
the $k$-th minimal model of an ${\cal N}$=2 SCFT with $k$ even \cite{BFK}.
This is very compelling evidence that each of these systems is quantum critical.

The fits become less reliable for levels with higher energies, but also if there
is an avoided level crossing as a function of the twist. This happens when the energy levels in the Ramond or
NS sector are degenerate. For the chain, the avoided crossings vanish in the continuum
limit \cite{FSd} (see also \cite{YF}), so one would expect that for the other models this is also merely a finite size
feature. For the
square ladder [$\vec{u}=(L,0)$ and $\vec{v}=(0,2)$], however, the results from exact
diagonalization suggest that the avoided
crossing will prevail for large system sizes. We are investigating
this issue using density matrix renormalization group methods \cite{HCS}.

\begin{figure}[h!]
\begin{center}
\includegraphics[width= .20\textwidth]{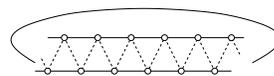}
\caption{The figure shows the zigzag ladder [$\vec{u}=(L,0)$ and $\vec{v}=(1,2)$], where the zigzagged
line is dashed to show that it can also be thought of as a chain with nearest \emph{and} next-nearest
neighbors excluded.}
\label{fig:zigzag}
\end{center}
\end{figure}

For the zigzag ladder [$\vec{u}=(L,0)$ and $\vec{v}=(1,2)$, see
fig. \ref{fig:zigzag}], we have computed the entanglement entropy
of the zero-energy ground states of the system up to 35 sites. We
find that for 1/4 and 1/5 filling, the entanglement entropy of the
ground states saturates. This implies that the correlation length
is finite, and thus the ground states at 1/4 and 1/5 filling are
not critical. However, at intermediate fillings we find good
correspondence to the spectral flow behavior of an ${\cal N}$=2
SCFT. As anticipated in the introduction, this indicates that the system is critical at 2/9 filling, where
there is a phase transition due to competing orders of the
diamonds and squares.

The systems we have investigated numerically are rather confined in one direction. This suggests that
we have essentially probed the edge modes and confirmed that they are gapless. Whether there are also
gapless modes in the bulk remains unclear from this analysis. However, for a more generic case such as
the triangular lattice, where the ground state entropy is truly extensive, one could speculate that a
similar argument as above would imply that there are critical modes in the bulk as well.

It will be most interesting to investigate whether or not
our main findings (quantum charge frustration and quantum 
criticality induced by strong repulsive interactions for
itinerant lattice fermions) carry over to more generic models,
and if they can be linked to some of the poorly understood 
physical features, transport properties in particular, of 
strange metals and heavy fermions systems displaying non-Fermi 
liquid behavior.

\acknowledgments{We thank Pasquale Calabrese for suggesting the entanglement entropy computation. K.S. and
L.H. acknowledge financial support through the Research Networking Programme INSTANS of the ESF. P.F. and J.H. 
were supported in part by NSF grants DMR-0412956 and DMR/MSPA-0704666.}

\end{document}